\documentclass[twocolumn,preprintnumbers,showpacs,amsmath,amssymb]{revtex4}
\usepackage{graphicx}% Include figure files
\usepackage{dcolumn}% Align table columns on decimal point
\usepackage{bm}% bold math

\begin{document}

\title{
 Hagedorn Thermostat: A Novel  View of Hadronic Thermodynamics}

%%%\author{L. G. Moretto$^{1}$,  K. A. Bugaev$^{1,2}$,  J. B. Elliott$^{1}$ and  L. Phair$^{1}$}
%
\author{ K. A. Bugaev$^{1,2}$, J. B. Elliott$^{1}$, L. G. Moretto$^{1}$  and  L. Phair$^{1}$}
\affiliation{
$^1$Lawrence Berkeley National Laboratory, Berkeley, CA 94720, USA\\
$^2$Bogolyubov Institute for Theoretical Physics,
Kiev, Ukraine
}

\date{\today}
\begin{abstract}
\noindent
A  microcanonical  treatment   of  Hagedorn systems, i.e. finite mass hadronic resonances with  an exponential mass spectrum controlled by the  Hagedorn temperature $T_H$,  is performed.
We show that, in  the absence of any restrictions,  a Hagedorn system is a perfect thermostat, i.e.
it imparts its temperature $T_H$ to any other system  in thermal contact with it.
We study the thermodynamic effects  of  the lower mass  cut-off in the Hagedorn mass spectrum.
We show  that in the  presence of a single
Hagedorn resonance   the temperature of any number of $N_B$  Boltzmann particles differs 
only slightly  from $T_H$ up to the kinematically allowed limit $N_B^{kin}$.
For $N_B > N_B^{kin}$ however, the low mass cut-off  leads to a decrease of the temperature as 
$N_B$ grows. 
The properties  of  Hagedorn  thermostats naturally  explain a  single value of hadronization temperature 
observed in elementary particle collisions at  high energies  and lead to some experimental predictions.
\end{abstract}

\pacs{25.75.-q,  25.75.Dw, 25.75.Nq, 13.85.-t}

\preprint{LBNL-57363}

\maketitle

%%%%%%%%%%%%%%%%%%%%%  Introduction

\section{Introduction}

\vspace*{-0.4cm}

The statistical bootstrap model (SBM) \cite{Hagedorn:65,Hagedorn:68}  gave the first evidence
that an  exponentially growing hadronic  mass spectrum
$g_H (m) =  \exp[ m/ T_H] ~ (m_{\rm o}/ m)^{a}$
for $m \rightarrow \infty$ (the constants $ m_{\rm o} $ and $a$ will be defined later) could lead to  new thermodynamics above the Hagedorn temperature
$T_H$. 
Originally, the divergence of thermodynamic functions at temperatures $T$  above $T_H$
was interpreted as the existence of a limiting temperature for hadrons.
In other words,
it is impossible
to build the hadronic thermostat above $T_H$.
A few years later an  exponential form of the  asymptotic mass spectrum  was found in
the MIT bag model \cite{MITBagM} and the associated limiting temperature was
interpreted as the phase transition temperature  to the partonic degrees of freedom \cite{Parisi:75}.
These results initiated  extensive studies of hadronic thermodynamics within the framework of
the gas of bags  model (GBM) \cite{Kapusta:81,Kapusta:82}.
The SBM with a non-zero  proper volume \cite{Vol:1}
of hadronic bags was solved analytically \cite{Gorenstein:81}
by the Laplace transform to the isobaric ensemble and the existence of phase transition
from hadronic to partonic matter (also called the  quark gluon plasma, QGP)
was shown.
Since then  this technique has been  used  to solve  more sophisticated
versions \cite{SBM:new,SBM:width} of the SBM  and  other statistical models
\cite{SMM:1}.

The major achievement of the SBM  is  that it naturally explains why the
temperature of secondary hadrons created in hadronic collisions cannot exceed $T_H$.
However,  this  result  is based  on two related  assumptions.
First, the grand canonical formulation for SBM is appropriate, and
second, the resonances of infinite mass should contribute
to thermodynamic functions.

Very recently, using the microcanonical  formulation,  we showed  \cite{Moretto:05}  that 
in the absence of any restrictions  on the mass,   resonances with the  Hagedorn 
mass spectrum behave as a perfect thermostat and perfect chemical reservoir,
i.e. they impart the Hagedorn temperature $T_H$  to  particles which are in thermal
contact and force them to be in chemical equilibrium. 
Similar questions about  chemical properties of the heavy resonances were addressed 
in \cite{Greiner:04}.

Our  analysis  \cite{Moretto:05} based on very general thermodynamic arguments  shows  that 
{\it it is improper to include 
any temperature other than $T_H$} 
into canonical and/or grand canonical formulations of the statistical mechanics 
of any system coupled to a Hagedorn thermostat. 
We also  demonstrated   that the Hagedorn thermostat  generates a volume independent 
concentration of the particles  in chemical equilibrium  with it \cite{Moretto:05}.
Thus, the entire  framework of the SBM
and GBM, which is also based on these two assumptions, must be revisited. 
In other words,   it is necessary to return to the foundations of the statistical mechanics of hadrons
and study the role of the  Hagedorn mass spectrum 
for  finite masses  of hadronic resonances above the cut-off value $m_{\rm o}$, 
below which the hadron mass spectrum  is discrete.  
Such an analysis for an arbitrary  value of  $a$  in  $g_H (m)$
was not done in   \cite{Moretto:05}, and, 
will be performed  here. 

This refinement  is important 
for  understanding  the differences and similarities
between  A+A and elementary particle 
collisions at high energies.
There are two temperatures measured in A+A collisions that are very close to
the transition temperature $T_{Tr}$ from hadron gas to QGP  calculated 
from  the lattice quantum 
chromodynamics  \cite{Lattice}  at vanishing baryonic density. 
The first  is the 
chemical freeze-out 
temperature
at vanishing baryonic density 
 $T_{Chem}  \approx 175 \pm 10$ MeV of the most  abundant hadrons (pions, kaons, nucleons 
{\it etc})  extracted from particle multiplicities  
at highest  SPS \cite{SPS:Chem} and all RHIC \cite{RHIC:Chem} energies.
Within the error bars $T_{Tr} \approx T_{Chem}$ 
is  also  very close to 
the kinetic freeze-out temperature  $T_{Kin} $ (i.e. hadronization temperature)
found from the transverse mass spectra 
of   heavy,  weakly interacting hadrons
such as
$\Omega$ hyperons, $J/\psi$ and $\psi^\prime$ mesons at the highest SPS  energy 
\cite{SPS:Hadronization1, SPS:Hadronization2}, 
and
$\Omega$-hyperons \cite{RHIC:Hadronization1, RHIC:Hadronization2, RHIC:Hadronization3} 
and $\phi$ meson \cite{RHIC:Hadronization1, RHIC:Hadronization3, RHIC:Hadronization4}    
at   $\sqrt{s} = 130$ A$\cdot$GeV and $\sqrt{s} = 200$ A$\cdot$GeV
energies of  RHIC.
The existence of the deconfinement  transition naturally explains the same value
for all these temperatures.

Can 
the same logic  be  applied to the collisions  of elementary particles, 
where  the  formation of a deconfined quark gluon matter 
 is rather problematic? 
The fact that 
the  hadronization  temperature \cite{Becattini:1} 
and inverse slopes of the transverse mass spectra  of  various 
hadrons \cite{Gazdzicki:04} found in  elementary particle  collisions at high energies 
are similar to those ones for A+A collisions
is tantalizing.

In the present paper we show that
the results of  A+A and elementary particle collisions can be understood and explained  on the same footing  by 
the Hagedorn thermostat concept. 
For this purpose 
we 
study  the properties of microcanonical  equilibrium  of  Boltzmann particles which are in contact
with  Hagedorn thermostats  and  elucidate the effect  of the mass cut-off
$m_{\rm o}$ of the Hagedorn spectrum
on the  temperature of  a system at  given energy. 
The microcanonical formulation of the Hagedorn thermostat model (HTM) is given in Section II.
Section III is devoted to the analysis of the most probable state of 
a  single heavy Hagedorn thermostat  of mass  $m \ge m_{\rm o}$ and  $N_B$  Boltzmann particles. 
The last section contains our conclusions and possible experimental consequences.

 %%%%%%%%%%%%%%%%% Section II
 
\vspace*{-0.3cm}

\section{Hagedorn Thermostat Model}

 \vspace*{-0.4cm}

Let us consider the microcanonical ensemble of $N_B$ Boltzmann point-like particles 
of mass $m_B$ and degeneracy $g_B$, 
and $N_H$  hadronic  point-like resonances of mass $m_H $ with a 
mass spectrum $g_H (m_H) = \exp[ m_H/ T_H ] (m_{\rm o}/ m_H )^a$ for $m_H \ge m_{\rm o}$
which obeys the inequalities
$m_{\rm o} \gg T_H$ and $m_{\rm o} > m_B$.
A recent analysis \cite{Bron:04}  suggests  that the  Hagedorn mass spectrum  can be established  for $m_{\rm o}  < 2$ GeV.

In the SBM \cite{Frautschi:71} and the  MIT bag model  \cite{Kapusta:81}
 it was found that for $m_H \rightarrow \infty$  the parameter $a \le 3$.
For finite resonance masses  the  value of $a$  is  unknown, so   it will be   considered as a fixed parameter.

The microcanonical partition 
of the system,  with  volume $V$,  total energy $U$ and zero total momentum,  can be written as follows 

\vspace*{-0.3cm}
\begin{equation*}
\Omega
%%%( U, N_H, N_B) 
= 
\frac{V^{N_H} }{ N_H !}  \left[ \prod_{k = 1}^{N_H}  g_H(m_H)\hspace*{-0.1cm} \int \hspace*{-0.1cm} 
\frac{ d^3 Q_k}{(2 \pi)^3 } \right] 
%
%%%&&\hspace*{-0.2cm}
\frac{V^{N_B} }{ N_B !}  \left[ \prod_{l = 1}^{N_B} g_B \hspace*{-0.1cm}\int  \hspace*{-0.1cm}
\frac{ d^3 p_l}{(2 \pi)^3 } \right] 
\end{equation*}
\begin{equation} \label{mone}
 \hspace*{0.4cm}
\delta\biggl( U - \sum\limits_{i = 1 }^{N_H} \epsilon^H_i - \sum\limits_{j = 1 }^{N_B} \epsilon^B_j \biggl) ,
\end{equation}
where the quantity $ \epsilon^H_i~=~\varepsilon(m_H, Q_i )$ $\left( \epsilon^B_j = \varepsilon(m_B, p_j) \right.$ and
$\left. \varepsilon(M, P) \equiv  \sqrt{M^2 + P^2}  \right)$  denotes  the energy  of
the Hagedorn (Boltzmann) particle with the 3-momentum ${\vec Q}_i$ (${\vec p}_j$).
In order to simplify the  presentation of our idea,  Eq. (\ref{mone}) accounts  for   energy conservation only 
and neglects  momentum conservation.

The microcanonical partition (\ref{mone}) can be evaluated by the Laplace transform 
in total energy $U$ \cite{Pathria}.
Then the momentum integrals in (\ref{mone}) are factorized and can be performed
analytically. The inverse Laplace transform in the conjugate variable $\lambda$ can be 
 done analytically for
 the nonrelativistic and ultrarelativistic approximations  of 
the one-particle momentum distribution  function
\begin{eqnarray}\label{mfour}
\hspace*{-0.cm} 
 \int\limits_0^\infty \hspace*{-0.1cm}  
\frac{d^3 p ~  {\textstyle e^{-\lambda \varepsilon(M , p ) } }   }{ ( 2 \pi)^3 } 
 \approx  
 \left\{
\begin{tabular}{ll}
\vspace{0.1cm} \hspace*{-0.1cm}$ \left[ \frac{ 2 M }{\lambda} \right]^{\frac{3}{2} }\hspace*{-0.1cm} I_{\frac{1}{2} }  e^{- M \lambda } \,,$
&  $  M  Re (\lambda)  >\hspace*{-0.1cm}> 1$\,, \\
\hspace*{-0.1cm}$ \frac{ 2 }{  \lambda^3 } ~ I_{ 2 } \,, $  & $  M  Re (\lambda)  < \hspace*{-0.1cm}< 1$\,.
\end{tabular}
\right.
%\nonumber
\end{eqnarray}
where the auxiliary integral is denoted as 
\begin{equation}\label{mfive}
I_b ~\equiv ~ \int\limits_0^\infty \hspace*{-0.0cm}  
\frac{d \xi }{ (2 \pi)^2 } ~ \xi^b
~{ e^{-\xi } }  \,.
\end{equation}
\vspace*{-0.30cm}

Since the formal steps of further evaluation  
are similar for  both cases, we discuss 
in detail the nonrelativistic limit only,  and later 
present  the results for the other case.  
The  nonrelativistic approximation ($  M  Re (\lambda)  >\hspace*{-0.1cm}> 1$) for  Eq.  (\ref{mone}) is as follows
\vspace*{-0.1cm}
\begin{eqnarray}
&&
\hspace*{-0.3cm} 
\Omega_{nr} 
%%%( U, N_H, N_B) 
= 
\frac{ \left[  V g_H(m_H)  \left[ 2 m_H  \right]^{ \frac{3}{2} }    I_{ \frac{1}{2} }  
\right]}{N_H!}^{N_H}   \nonumber
\\
%
%\nonumber \\
%%%%%\end{eqnarray}
%
%%%%%\vspace*{-0.5cm}
%
%%%%%\begin{eqnarray}
%
\label{meight}
&&
\hspace*{0.8cm} 
 \frac{ \left[  V g_B \left[ 2 m_B \right]^{ \frac{3}{2} }  ~ I_{ \frac{1}{2} } \right] }{N_B!}^{N_B} 
\hspace*{-0.2cm} 
\frac{ E_{kin}^{\frac{3}{2} (N_H + N_B) - 1}  }{  \left( \frac{3}{2} (N_H + N_B) - 1 \right)! }\,,
\end{eqnarray}
where $E_{kin} = U - m_H N_H - m_B N_B $ is the kinetic energy of the system.

As shown below, the most realistic case 
corresponds to the nonrelativistic treatment of the Hagedorn resonances because 
the resulting temperature is  much smaller than 
their masses. Therefore, it is sufficient to consider the ultrarelativistic
limit for the Boltzmann particles only. In this case 
 ($  M  Re (\lambda)  <\hspace*{-0.1cm}< 1$)
 the equation (\ref{mone}) can be approximated as 
\begin{eqnarray}\label{mten}
&&\hspace*{-0.6cm} \Omega_{ur} 
%%%( U, N_H, N_B)  
= 
\frac{ \left[  V g_H(m_H)  \left[ 2 m_H  \right]^{ \frac{3}{2} }   ~ I_{ \frac{1}{2} }  
\right]}{N_H!}^{N_H}   
\nonumber \\
&&\hspace*{0.5cm} 
 \frac{ \left[  V g_B  ~2  ~ I_{ 2 } \right] }{N_B!}^{N_B} 
\hspace*{-0.2cm} 
\frac{ E_{kin}^{\frac{3}{2} (N_H + 2 N_B) - 1}  }{  \left( \frac{3}{2} (N_H + 2 N_B) - 1 \right)! }\,,
\end{eqnarray}
where the  kinetic energy does not include the rest energy  of the Boltzmann particles, i.e.
$E_{kin} = U - m_H N_H $.

Within our assumptions 
 the above results are general
and can be used for any number of particles,  provided $N_H + N_B \ge 2$.
It is instructive to consider first  the  simplest case $N_H = 1$.  
This oversimplified model, in which a Hagedorn thermostat is always present,  allows
us to study the problem rigorously.
For  $N_H = 1$ we treat  the mass of Hagedorn
thermostat $m_H$  as a free parameter and determine the  value which maximizes the entropy of the system.  The 
solution $ m_H^* > 0$  of 
%%%the  extremum condition 
%
\begin{eqnarray}\label{meleven}
\hspace*{-0.5cm}&& \frac{ \delta \ln \Omega_{nr} (N_H = 1) }{\delta~ m_H } ~ =  \nonumber \\
\hspace*{-0.5cm}&& {\textstyle  \frac{1}{T_H}~ +~ \left( \frac{3}{2}~ - ~ a \right) \frac{1}{m_H^*} ~ - ~  
\frac{3 (N_B + 1) }{2~ E_{kin} } ~ = ~ 0 } 
\end{eqnarray}
provides the maximum of the system's entropy, if  for $m_H = m_H^*$ the second derivative is negative
\begin{eqnarray}\label{mtwelve}
\hspace*{-0.5cm}&& \frac{ \delta^2 \ln \Omega_{nr} (N_H = 1) }{\delta~ m_H^2 } ~ =  \nonumber \\
\hspace*{-0.5cm}&& {\textstyle - ~ \left( \frac{3}{2}~ - ~ a \right) \frac{1}{m_H^{*\,2} } ~ - ~  \frac{3 (N_B + 1) }{2~ E_{kin}^2 } ~ < ~ 0 \,. }
\end{eqnarray}
If  the inequality  (\ref{mtwelve}) is satisfied, then the extremum condition (\ref{meleven})
defines the  temperature of the system of $(N_B + 1)$ nonrelativistic particles
\begin{equation}\label{mthirteen}
\hspace*{-0.25cm} T^* (m_H^*)  \equiv  
\frac{ 2 ~ E_{kin} }{  3 (N_B + 1) } = \frac{T_H}{ 1 ~ + ~ \left(  \frac{3}{2}~ - ~ a \right) \frac{T_H }{m_H^{*} }    }  \,.
\end{equation}
Thus, as $m_H^* \rightarrow \infty$ it follows that $T^*(m_H^*) \rightarrow T_H$, while for 
finite $m_H^* \gg T_H$  and $ a >  \frac{3}{2} $  ($ a <  \frac{3}{2} $) the temperature 
of the system is 
slightly larger  (smaller) than the Hagedorn temperature, i.e. $ T^* > T_H$ ($ T^* < T_H$).
Formally, the temperature of the system in equation (\ref{mthirteen}) may differ  essentially 
from $T_H$ for  a  light thermostat, i.e. for $m_H^* \le T_H$.
However, it is assumed that 
the Hagedorn mass spectrum exists
above the cut-off mass $m_{\rm o} \gg T_H$, thus $m^* \gg T_H$.

%%%%%%%%%%%%%%%%%%%%% Section III

\vspace*{-0.3cm}

\section{The Role of the Mass Cut-off}

\vspace*{-0.3cm}

Now we study the effect of the mass cut-off of the Hagedorn spectrum on the
inequality (\ref{mtwelve}) in more detail. 
For $ a \le  \frac{3}{2} $ the condition (\ref{mtwelve}) is  satisfied. For $ a >   \frac{3}{2} $ the inequality (\ref{mtwelve})
is equivalent to 
\begin{equation}\label{mfourteen}
\hspace*{-0.25cm}
 \frac{ m_H^{*\,2}  }{   \left( a -  \frac{3}{2} \right)  ~ T^*(m_H^*)    }  ~ > ~ 
\frac{ 3 }{ 2 } ~  (N_B + 1) ~ T^*(m_H^*)  
\,,
\end{equation}
which  means that a  Hagedorn thermostat should be massive  compared to the kinetic energy of the system.  

A more careful analysis shows that 
for a negative value of the determinant $ D_{nr} $  $( \tilde{N} \equiv N_B - \frac{2}{3} a )$  
\begin{eqnarray}\label{mfifteen}
\hspace*{-0.25cm}
D_{nr} & \equiv & {\textstyle  \left( U - m_B N_B - \frac{3}{2}~ T_H ~ \tilde{N} \right)^2 - } \nonumber \\ 
& & {\textstyle 4 \left( a - \frac{3}{2} \right)~T_H~ \left( U - m_B N_B \right) ~ < ~ 0\,, }
\end{eqnarray}
equation (\ref{meleven}) has two complex solutions, while  for $D_{nr}~=~ 0$ there exists a single 
real solution of (\ref{meleven}).
Solving (\ref{mfifteen}) for $(U - m_B N_B)$, shows that  for $\tilde{N} > \frac{2}{3} a - 1$,
 i.e. for ${N_B} > \frac{4}{3},  a - 1$ the inequality (\ref{mfifteen})  does not hold and $D_{nr} > 0$.
Therefore, in what follows we will assume that ${N_B} > \frac{4}{3} a - 1$ and  only analyze  the case $D_{nr} > 0$. 
For this case  equation (\ref{meleven}) has two real solutions 
\begin{equation}\label{msixteen}
m_H^\pm = {\textstyle \frac{1}{2} \left[ U - m_B N_B - \frac{3}{2}~ T_H ~ \tilde{N}~ \pm ~ \sqrt{ D_{nr} } \right]\,.} 
\end{equation}
For $ a \le  \frac{3}{2} $ only $m_H^+$ solution is positive and
corresponds to a maximum of  the microcanonical partition $\Omega_{nr}$. 

For $ a >  \frac{3}{2} $ both solutions of (\ref{meleven}) are positive, but only  $m_H^+$  is a maximum.
From the two limiting cases:
\begin{eqnarray}\label{mseventeen}
\hspace*{-0.5cm}
\frac{ \delta \ln \Omega_{nr} (N_H = 1) }{\delta~ m_H } & \approx &  
{\textstyle \left( \frac{3}{2} -  a \right) \frac{1}{m_H}  \quad {\rm for} \quad m_H \approx 0\,,} 
\\
\label{meighteen}
\hspace*{-0.5cm}
%
%{\textstyle
\frac{ \delta \ln  \Omega_{nr} (N_H = 1)  }{\delta~ m_H }  & \approx &
{\textstyle 
\frac{3 (N_B + 1) }{2~ E_{kin} } \quad {\rm for} \quad E_{kin} \approx 0\,,}
\end{eqnarray}
and the fact that $ m_H^\pm $ 
obey the   inequalities
\begin{equation}\label{mnineteen}
0 ~ < ~  m_H^- ~  \le ~ m_H^+ ~ < ~ U - m_B N_B \,, 
\end{equation}
it is clear that $ m_H^* = m_H^-$ is a local minimum  of the microcanonical partition $\Omega_{nr}$,
while  $ m_H^* = m_H^+$ is  a local maximum of the partition $\Omega_{nr}$.

Using Eq. (\ref{msixteen}) for $m_H^+$, it is clear that 
for any value of  $a$ 
the constraint $m_H^+ \ge m_{\rm o} $
is equivalent to the  inequality 
\begin{equation}\label{mtwenty}
N_B ~ \le ~ N_B^{kin} \equiv  { \frac{ U ~ - ~ [ \frac{m_{\rm o} }{T_H}~ - ~a ] ~ T^*(m_{\rm o})  }{ m_B \, + \, \frac{3}{2} ~ T^*(m_{\rm o})  }  } \,.
\end{equation}
Thus, at fixed energy $U$   for all  $N_B \le N_B^{kin}$ 
 at  $m_H^*  = m_H^+$ there is a local maximum of the microcanonical partition  $\Omega_{nr}$ with
the temperature $ T = T^*( m_H^+) $. For $N_B > N_B^{kin}$   the maximum of the partition $\Omega_{nr}$ cannot be reached due to the cut-off constraint
and, consequently,   the most probable state corresponds to $m_H = m_{\rm o}$
with 
$T  \le T^*(m_{\rm o} )$ from Eq. (\ref{mthirteen}). 
In other words, for $N_B > N_B^{kin}$  the amount of energy $U$ is insufficient for  the mass of the Hagedorn thermostat to be above the cut-off  $m_{\rm o} $ 
and   simultaneously  maintain  the  temperature of the Boltzmann particles according to  
Eq. (\ref{mthirteen}).  
By assumption there is a single Hagedorn thermostat in the system, therefore,  
as $N_B$ grows the temperature of the system decreases
from $T^*(m_{\rm o} )$ value.
Thus, the equality (\ref{mtwenty}) defines the kinematical limit 
for reaching the maximum of the microcanonical partition.

To prove that the maximum of the microcanonical 
partition at $m_H = m_H^+$  is  
global  it is sufficient to show that 
the  constraint $m_H^+ \ge m_{\rm o}$
is not consistent with the condition $m_H^- > m_{\rm o}$. 
For $a \le \frac{3}{2}$ the maximum is  global  because for $ 0 <  m_H < m_H^+$ 
($m_H >  m_H^+$ ) the partition $\Omega_{nr} (N_H =1, m_H ) $ monotonically increases (decreases)
with $m_H$. 
For $a > \frac{3}{2}$  it is clear that the maximum at $m_H = m_H^+$ is local, 
if  the state with mass  $m_H =  m_{\rm o}$ is more probable, i.e. 
$\Omega_{nr} (N_H =1, m_{\rm o}) >  \Omega_{nr} (N_H =1, m_H^+ ) $.  Due to (\ref{mnineteen})  this can occur, if $m_H^- > m_{\rm o}$.  Substituting Eq.  (\ref{msixteen})  into the last inequality,
shows  that this inequality
reduces  to the condition $N_B > N_B^{kin}$.
This contradicts  the  constraint $m_H^+ \ge m_{\rm o}$ in the form of  Eq. 
(\ref{mtwenty}). 
Thus, the maximum of the
microcanonical partition is  global.

To complete our consideration of the  nonrelativistic case 
let us express the partition (\ref{meight}) in terms
of the temperature (\ref{mthirteen}).  Applying the Stirling approximation  to  the  factorial $(\frac{3}{2}(N_B + 1) -1 )!$
for 
$N_B^{kin} > N_B \gg 1$ and reversing the
integral representations (\ref{mfour})  and (\ref{mfive}) for $\lambda = 1/ T^*(m_H^+)$, one
finds 
\begin{eqnarray}\label{mtwone}
&&\hspace*{-0.6cm} \Omega_{nr} (N_H=1) ~ = ~  
%
%%%{e^{\frac{U}{ T^*(m_H^+) } } } ~
%
\frac{  V \, g_H(m_H^+)  }{  T^*(m_H^+) } ~  \hspace*{-0.0cm}  \int  \hspace*{-0.0cm}  
\frac{d^3 Q}{ (2 \pi)^3 } 
~{\textstyle e^{ - \frac{  \sqrt{m^{+\,2}_H + Q^2}   }{  T^*(m_H^+) }  } }   \nonumber \\
&&\hspace*{-0.6cm}  \frac{ e^{\frac{U}{ T^*(m_H^+) } }  }{N_B!}  \left[ V \, g_B  \hspace*{-0.0cm}  \int  \hspace*{-0.0cm}  
\frac{d^3 p}{ (2 \pi)^3 } 
~{\textstyle e^{ - \frac{  \sqrt{m^2_B + p^2}   }{  T^*(m_H^+) }  } }  \right]^{N_B}  \,.
\end{eqnarray}
This is just 
the  grand canonical partition of $(N_B + 1)$ Boltzmann
particles with temperature $ T^*(m_H^+) $. 
If $N_B > N_B^{kin} \gg 1$, then $ T^*(m_H^+)$
in (\ref{mtwone}) should be 
replaced  by 
$T_{\rm o} (N_B) \equiv  \frac{2(U - m_B N_B - m_{\rm o} ) }{3 (N_B + 1) } $. 

Fig. 1 shows that 
for  $a >  \frac{3}{2}  $
the system's temperature $T = T^*(m_H^+)$  as a function of $N_B$ remains almost  constant    for 
 $N_B < N_B^{kin}$, reaches a maximum  at  $N_B^{kin} $  and rapidly decreases like 
 $T  = T_{\rm o}  (N_B) $  for  $N_B > N_B^{kin} $.
For $a <  \frac{3}{2} $ the temperature  has a plateau $T =  T^*(m_H^+)$ for $N_B <  N_B^{kin}$,
and rapidly    decreases  for 
$N_B > N_B^{kin} $ according to  $T_{\rm o}  (N_B) $. 

The same results  are  valid for the ultrarelativistic treatment of Boltzmann 
particles.  Comparing the nonrelativistic and ultrarelativistics expressions for
the microcanonical partition, i.e. equations (\ref{meight}) and (\ref{mten}), respectively,
one  finds that   the derivation of the  ultrarelativistic limit   requires only 
the substitution $N_B \rightarrow 2 N_B$ and $m_B / T_H \rightarrow 0$
in  equations (\ref{meleven} -- \ref{mtwone}). 
Note that  this  substitution does not alter  the expression for the temperature
of the system, i.e. the right hand side of (\ref{mthirteen}). 

%%%The last step of out study is to generalize the above results to any number of Hagedorn %%%thermostats and any number of Boltzmann particles. This can be done straightforwardly. 

Finally,  we show that for a heavy Hagedorn thermostat  ($m_H^+  \gg  m_{\rm o}$) these results remain 
valid for  a single Hagedorn thermostat split  into $N_H$ pieces of the same mass.
 %%% being heavier than $ m_{\rm o} $.  
Substituting $m_H \rightarrow m_H N_H$ in the nonrelativistic expressions 
(\ref{meight}) and minimizing  it with respect to $m_H$,  the temperature of the system in the form of  equation (\ref{mthirteen})  is $ T^*(m_H^* N_H) $, where the mass of $N_H$ Hagedorn thermostats $m_H^*$ is related to the solution $m_H^+ $ of equation (\ref{msixteen})   as  
$m_H^* = m_H^+ / N_H$. Since the original single thermostat of mass $m_H^+ $ was assumed to be heavy,  it follows $T^*(m_H^* N_H) = T^*(m_H^+) \rightarrow T_H$. 
A more careful study (see also   \cite{Moretto:05})  using
an exact expression for the microcanonical partition of $N_H$ Hagedorn thermostats of the same mass
$m_H$ gives the same result, if  $m_H  \gg  m_{\rm o} $. 
A generalization of these statements to  the case of $N_H$ heavy 
Hagedorn thermostats of different masses  also leads to the same result.  
Thus,  splitting  a single
heavy Hagedorn thermostat into an arbitrary number of heavy resonances (heavier than $ m_{\rm o} $) does not
change the temperature of the system.

%%%%%%%%%%%%%%%%%%%%%%%%%% Figure 2
%
%
\begin{figure}[ht]
\includegraphics[width=8.6cm,height=8.6cm]{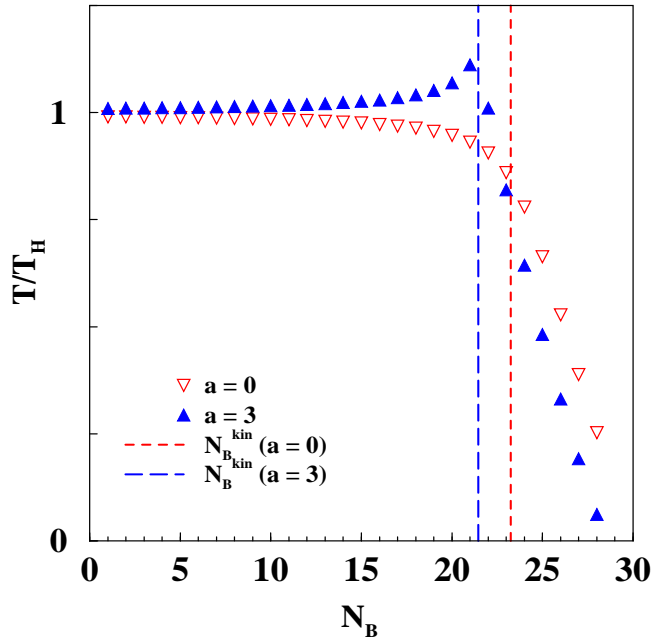}
\vspace*{-0.5cm}
\caption{
A typical behavior  of the system's temperature  as the function of the  number 
of Boltzmann particles $N_B$ for $ a = 3$ and $a = 0$ for the same value of the total energy 
$U = 30 \,m_B$.
Due to the thermostatic properties of a Hagedorn resonance
the system's temperature  is nearly constant 
up to the kinematically allowed value $N_B^{kin}$ given by (\ref{mtwenty}). 
}
  \label{fig2}
\end{figure}

\vspace*{-0.3cm}

%%%%%%%%%%%%%%%%%%%%% Section IV

\vspace*{-0.3cm}

\section{Conclusions}

\vspace*{-0.3cm}

In the present work we generalized the  SBM  results \cite{Frautschi:71}  to  systems  of finite energy by showing  explicitly
that  even  a single resonance with  the Hagedorn mass spectrum degeneracy,
i.e. {\it a Hagedorn thermostat,}  keeps an almost constant   
temperature  close to $T_H$  for  any  number of Boltzmann particles $3 < N_B  \le N_B^{kin} $. 
For the  high energy limit  $U \gg m_{\rm o}$ this means that 
a single  Hagedorn resonance defines 
the temperature of the system to be only slightly  different from $T_H$ until 
 the energy of the Hagedorn thermostat is almost negligible compared to $U$.
In contrast to the grand canonical formulation of the original SBM  \cite{Frautschi:71},
in the presence of a Hagedorn thermostat 
the  temperature $T_H$ 
can be reached  at  any energy density.

The thermostatic  nature of a Hagedorn system  
obviously
 explains the ubiquity of both the inverse slopes of  measured transverse mass spectra 
\cite{Gazdzicki:04} and hadronization temperature found in  numerical simulations 
of hadrons
created in elementary particle collisions at high  energies \cite{Becattini:Can,Becattini:1, Becattini:2}.
 By a direct evaluation of the microcanonical partition
we showed that  in the presence of a single Hagedorn thermostat  the energy spectra of particles become 
exponential 
 with no  additional assumptions, e.g. {\it  phase space dominance} \cite{PSDominance} or
{\it string tension fluctuations} \cite{Strings}.
Also the limiting temperature found in the URQMD calculations made in a  finite box 
\cite{URQMD:Box} can be explained by the effect of the Hagedorn thermostat. 
We expect that, if the string parametrization of the URQMD in a  box \cite{URQMD:Box} was done microcanonically instead of grand canonically,  then  the same behavior  would be found.

The Hagedorn thermostat model
generalizes the statistical hadronization model which successfully
describes the particle multiplicities in nucleus-nucleus and elementary collisions \cite{Becattini:Can,Becattini:1, Becattini:2}.
The statistical hadronization model  accounts for   the decay of heavy resonances 
(clusters in terms of Refs. \cite{Becattini:Can,Becattini:1, Becattini:2}) only  and does not consider the additional
particles, e.g. light hadrons, free quarks and gluons, or other heavy resonances.
As we showed,
the splitting of a single
heavy Hagedorn resonance into several  does not
change the temperature of the system. 
This finding  justifies the main assumption of 
the canonical formulation of the statistical hadronization model \cite{Becattini:Can} that smaller clusters   
may be reduced 
to a single large cluster. 
Thus,  
recalling the  MIT Bag model interpretation of 
the Hagedorn mass spectrum  \cite{Kapusta:81, Kapusta:82}, we conclude that 
 quark gluon matter confined in  heavy resonances (hadronic bags)  
controls the  temperature of surrounding particles close to $T_H$, and, therefore,
this temperature can  be considered as 
a coexistence  temperature for  confined color matter and hadrons. 
Moreover, as we showed, the emergence  of a coexistence temperature does
not require the actual deconfinement of the color degrees of freedom,
which, in terms of the GBM \cite{Gorenstein:81}, is equivalent to 
the formation of the infinitely large
and infinitely heavy hadronic bag.

Within the framework of  the Hagedorn thermostat model we found that even for a single Hagedorn thermostat and 
 $a >  \frac{3}{2}  $
the system's temperature $T = T^*(m_H^+)$  as a function of $N_B$ remains almost  constant    for 
 $N_B < N_B^{kin}$, reaches a maximum  at  $N_B^{kin} $  and rapidly decreases  
 for  $N_B > N_B^{kin} $ (see Fig. 1). 
For $a <  \frac{3}{2} $ the temperature  has a plateau $T =  T^*(m_H^+)$ for $N_B <  N_B^{kin}$,
and rapidly    decreases  for 
$N_B > N_B^{kin} $.
If such characteristic behavior 
of the hadronization temperature or the hadronic inverse slopes 
%%%produced in events
%%%of different multiplicities  can be established experimentally, it   may  fix the value of  $a$.  
can be measured as a function of event multiplicity, it may be possible to experimentally
determine  the value of  $a$. 
For  quantitative predictions
it is necessary to include more hadronic species into the model, but this will not change 
our  result.

If we apply the HTM to elementary particle collisions at high energy, then, as  shown above, 
the temperature of  created particles will  be  defined by the
most probable mass of the Hagedorn thermostat. 
If the most probable resonance mass 
grows  with  the energy of collision,  then the hadronization  temperature should decrease (increase)
to $T_H$ for $a >  \frac{3}{2}  $ $(a <  \frac{3}{2}  )$. 
Such a decrease is  observed   in 
reactions of elementary particles  at  high energies, see Table 1 in Ref.
\cite{Becattini:2}.

In order to apply 
these results in a more physical fashion to the quark gluon plasma formation in relativistic 
nucleus-nucleus collisions (where
the excluded volume effects  are known to be important  
\cite{Vol:1, Gorenstein:81, SPS:Chem, SPS:Chem2}  
for all hadrons) 
 the eigen volumes of all particles  should be incorporated into the model.
For pions this should be done in relativistic fashion \cite{Bugaev:00a}.
Also the effect of finite  width of Hagedorn resonances may be important \cite{SBM:width}  and should  be studied.

%%%KKK

%For further evaluation it is convenient to  use the following exact representation of the modified Bessel function
%\cite{Gradstein}
%
%\begin{equation}\label{mfour}
%
%
 %K_2 ( M \lambda )  = \frac{ 4 M^2 ~e^{-\lambda M} }{3 ~\sqrt{2 \lambda} }
 %
 %\int\limits_0^\infty \hspace*{-0.1cm}  d t ~ e^{- M t} ~ t^{\frac{3}{2} } \left( 1 + \frac{t}{2 \lambda} \right)^{\frac{3}{2} } \,,
%
%\end{equation}
%

{\bf  Acknowledgments.}
The authors appreciate a lot the fruitful and stimulating discussions 
with V. Koch.
This work was supported by the US Department of Energy.

%%%%%%%%%%%%%%%%%%%%%%%%%%%%%%%%%%%%%%%%%%%%%%%%%%%%%%%%%%%%%%%%%%%%%
%%%\begin{references}
%%%\input surf_ref.tex
%%%\end{references}

\end{document}